\documentclass[sn-mathphys-num]{sn-jnl}% Math and Physical Sciences Numbered Reference Style 
\usepackage{graphicx}%
\usepackage{multirow}%
\usepackage{amsmath,amssymb,amsfonts}%
\usepackage{amsthm}%
\usepackage{mathrsfs}%
\usepackage[title]{appendix}%
\usepackage{xcolor}%
\usepackage{textcomp}%
\usepackage{manyfoot}%
\usepackage{booktabs}%
\usepackage{algorithm}%
\usepackage{algorithmicx}%
\usepackage{algpseudocode}%
\usepackage{listings}%
\theoremstyle{thmstyleone}%
\theoremstyle{thmstyletwo}%

\theoremstyle{thmstylethree}%

\begin{document}

\title[Article Title]{Lighthill's theory of sound generation in a non-isothermal 
medium}

%%=============================================================%%
%% GivenName	-> \fnm{Joergen W.}
%% Particle	-> \spfx{van der} -> surname prefix
%% FamilyName	-> \sur{Ploeg}
%% Suffix	-> \sfx{IV}
%% \author*[1,2]{\fnm{Joergen W.} \spfx{van der} \sur{Ploeg} 
%%  \sfx{IV}}\email{iauthor@gmail.com}
%%=============================================================%%

\author{\fnm{S.} \sur{Routh$^1$}}\email{swatirouth@dubai.bits-pilani.ac.in}
\author{\fnm{Z. E.} \sur{Musielak$^2$}}\email{zmusielak@uta.edu}

\affil{\orgdiv{$^1$General Science Department}, \orgname{BITS Pilani Dubai Campus}, 
\city{Dubai}, \postcode{ 345055}, \state{Dubai}, \country{UAE}}

\affil{\orgdiv{$^2$Physics Department}, \orgname{University of Texas at Arlington}, 
\city{Arlington}, \postcode{76019}, \state{Texas}, \country{USA}}

%%==================================%%
%% Sample for unstructured abstract %%
%%==================================%%

\abstract{Lighthill's theory of sound generation was developed to calculate 
acoustic radiation from a narrow region of turbulent flow embedded 
in an infinite homogeneous fluid. The theory is extended to include 
a simple model of non-isothermal medium that allows finding analytical 
solutions. The effects of one specific temperature gradient on the 
wave generation and propagation are studied. It is shown that that 
presence of the temperature gradient in the region of wave generation 
leads to monopole and dipole sources of acoustic emission, and that 
the efficiency of these two sources may be higher than Lighthill's
quadrupoles. In addition, the wave propagation far from the source is 
different than in Lighthill's original work because of the presence 
of the acoustic cutoff frequency resulting from the temperature 
gradient.}

%\keywords{}

%%\pacs[JEL Classification]{D8, H51}

%%\pacs[MSC Classification]{35A01, 65L10, 65L12, 65L20, 65L70}

\maketitle

\section{Introduction}

A theory of acoustic wave generation by a turbulent jet embedded in 
an infinite homogeneous fluid was originally developed by Lighthill [1] 
who showed that Reynolds stresses are sources of quadrupole emission.  
The theory allows evaluating the wave energy flux far away from a finite 
region of turbulence by assuming that the backreaction of generated 
waves on the turbulence is negligible.  An important result was obtained 
by Proudman [2] who described the turbulent motions in a jet by the 
Heisenberg turbulence energy spectrum [3], and derived a general 
formula for the generated acoustic power output.

The original Lighthill theory was also extended to include the effects 
of solid boundaries (e.g., [4,5]) and magnetic fields (e.g., [6,7]).  
The main prediction of the theory is the now well-known $u^8$ 
law of the acoustic power output by the turbulent jet, where $u$ 
is the jet velocity.  Good agreements between this theoretical 
prediction and the results of several experiments performed for 
jets of different diameters were reported by Goldstein [8].  The 
Lighthill-Proudman formula for the acoustic power output was 
used to evaluate the acoustic wave energy fluxes generated by 
turbulent motions in the solar (e.g., [9-11]) and stellar (e.g., 
[12,13]) convection zones, and to discuss the role played by 
acoustic waves in heating the solar and stellar atmospheres.

As mentioned above, the original Lighthill theory concerns 
only homogeneous media and treats turbulence as isotropic, 
homogeneous and decaying in time.  A significant extension 
of Lighthill's theory was done by Stein [14], who followed 
earlier work [15,16], and included the effects of stratification. 
Stein's treatment of turbulence was further improved and 
modfied in [16], and the resulting theory demonstrated that 
stratification is responsible for monopole and dipole sources 
of acoustic emission.  The modified Lighthill-Stein theory 
was used to compute the acoustic wave energy spectra for 
the Sun [16] and late-type stars [17].  Goldreich and Kumar 
[18] studied differences between free (decaying) and forced 
(non-decaying) turbulence, and demonstrated that the latter 
is driven by the fluctuating buoyancy force, which leads to 
dipole emission.  However, this resulting dipole emission 
shows similarities to that obtained by Stein [14]. 

In all the above applications of the theory of sound generation 
by turbulent motions, the background medium was assumed 
to be isothermal.  Therefore, the main aim of this paper is to 
extend the original Lighthill theory to a non-isothermal medium. 
The model of this medium is assumed to be simple, namely, its
density and temperature vary with height and they are related 
to each other, but pressure remains constant, which means that 
there is no gravity; for this model, analytical solutions can be 
obtained.  The solutions are then used to study the effects 
caused by one specific temperature gradient on the wave 
generation and propagation.  The obtained results show 
that the gradient leads to monopole and dipole sources 
of acoustic emission, and that it also is responsible for the 
acoustic cutoff frequency, which affects the wave propagation. 

The paper is organized as follows: a brief description of original 
Lighthill's theory is given in Sect. 2; extension of Lighthill's theory 
to a non-isothermal medium is described in Sect. 3; the acoustic 
wave energy fluxes computed by using both the original and 
extended Lighthill's theories are presented and compared in 
Sect. 4; and our conclusions are given in Sect. 5.

\section{Lighthill theory of sound generation} 

To describe the sound generation by a turbulent jet, Lighthill 
[1] derived an inhomogeneous wave equation for a single 
wave variable by collecting all linear and nonlinear terms on 
the left-hand side (the propagator) and on the right-hand side 
(the source function) of the wave equation, respectively, and
obtained
\begin{equation}
\hat L_s [ \rho ] = \hat S [ T_{ij} (u_t) ]\ ,
\label{eq:LWE}
\end{equation}
\noindent
where $\hat L_s$ is the acoustic wave propagator given by
\begin{equation}
\hat L_s = {{\partial^2} \over {\partial t^2}} - c_s^2 \nabla^2\ ,
\label{eq:LWO}
\end{equation}
and $\rho$ represents density perturbations associated with the 
waves, $c_s$ is the speed of sound, $u_t$ is the turbulent velocity,
and $T_{ij} (u_t) = \rho_o u_{ti} u_{tj} + p_{ij} - c_s^2 \rho 
\delta_{ij}$ is Lighthill's turbulence stress tensor with i,j = 1,2 
and 3, and $\rho_o$ being the density of the background medium. 
Lighthill assumed that the jet was embedded in an uniform 
atmosphere, which was also at rest, and considered linear (weak) 
acoustic waves that produce no backreaction on the turbulent flow.
He then showed that $T_{ij} \approx \rho_o u_{ti} u_{tj}$ and that 
the source function $\hat S [T_{ij} (u_t)]$ was given by a double 
divergence of $T_{ij}$. The physical meaning of the source function 
is that the stresses produce equal and opposite forces on opposite 
sides of a fluid element leading to the distortion of its surface 
without changing the volume (quadrupole emission). In other words, 
the fluid motions generating acoustic waves behave as a volume 
distribution of acoustic quadrupoles, so one may write $\hat S [ 
T_{ij} (u_t) ] = S_{quadrupole}$.

Proudman [2] applied Lighthill's theory to the case when the 
fluctuating fluid motions are represented by the Heisenberg 
turbulence energy spectrum [3] and derived a general 
formula for the generated acoustic power output, $P_a$.  
This Lighthill-Proudman formula is usually given in the 
following form:
\begin{equation} 
P_a = \alpha_q {{\rho_o u_t^3} \over {l_o}} M_t^5\ , 
\label{eq:LPF}
\end{equation}
where the emissivity coefficient $\alpha_q \approx 38$, $l_o$ 
is the characteristic length scale of the turbulence and 
$M_t = u_t / c_s$ is the turbulent Mach number. 

The formula is valid for subsonic turbulence ($M_t << 1$) 
and it was extensively used in early calculations of acoustic 
wave energy fluxes generated in the Sun and other stars 
(e.g., [9-13]).  Since the formula does not account for 
temperature gradients, it was assumed that Eq. (\ref{eq:LPF}) 
was satisfied locally in the turbulent region and the total emitted 
wave energy flux was calculated by performing the integration 
over the thickness of the wave generation region.

\section{Extension of Lighthill's theory}

\subsection{Basic equations}

Let us consider a compact region of turbulent flow embedded in 
a very large volume of an ideal gas and assume that both the 
turbulent region and the surrounding medium are non-uniform
because of the existence of a temperature gradient. To simplify 
the problem so that analytical solutions can be obtained, we 
neglect gravity and consider a model in which the gas pressure
$p_0$ = const, however, the background temperature $T_0$, 
density $\rho_0$ and speed of sound $c_s$ vary with height 
in the model.  The gradients of density and temperature that 
are related to each other preserve the hydrostatic equilibrium 
of the background medium. 

To describe the generation and propagation of acoustic waves 
in this model, we consider a set of hydrodynamic equations and 
assume that the turbulent flow is subsonic and the waves are 
linear. In general, the waves propagate in all three (x, y and z) 
directions, however, their propagation in the z-direction is 
affected by the gradient. We introduce $x_i$, with i = 1, 2 and 3, 
and define $x_1 = x$, $x_2 = y$ and $x_3 = z$. The waves are 
described by using the velocity $u_i (t, x_i)$, density $\rho 
(t, x_i)$ and pressure $p (t, x_i)$ perturbations. We further 
assume that the effects of viscosity and heat conduction can 
be neglected.  Based on these assumptions, we linearize the 
hydrodynamic equations and follow Lighthill [1] to separate 
the linear and nonlinear terms.  This gives
\footnote{We have used subscript notation for the Cartesian 
components of vectors and tensors, and any substript repeated 
in a single term is to be summed from 1 to 3.} 
\begin{equation} 
{\partial \rho \over {\partial t}} + {{\partial (\rho_0 u_i)}  
\over {\partial x_i}} = - {{\partial (\rho u_i)} \over {\partial 
x_i}}\ ,
\label{eq:continuity}
\end{equation}
\begin{equation}
\rho_0 {{\partial u_i}  \over {\partial t}} + {\partial p \over 
\partial x_i} = - {{\partial \rho u_i}  \over {\partial t}} - {{
\partial (\rho_0 u_i u_j)} \over {\partial x_j}}\ ,
\label{eq:momentum}
\end{equation}
and
\begin{equation} 
{{\partial p}  \over {\partial t}} + \rho_0 c_s^2 {\partial u_i 
\over \partial x_i} = - u_i {{\partial p}  \over {\partial x_i}} 
- c_s^2 \rho {{\partial u_i} \over {\partial x_i}}\ ,
\label{eq:energy}
\end{equation}
\\
\noindent
where, in general, $c_s = c_s (x_i)$ in our non-isothermal model.

\subsection{Wave equation and source function}

We derive an inhomogeneous wave equation for the pressure perturbation 
$p$ associated with the waves by eliminating the other wave variables, 
and obtain 
\begin{equation}
{\partial^2 p \over \partial t^2} - {\partial \over \partial x_i} 
\left [ c_s^2 (x_i)\ {\partial p \over \partial x_i} \right ]\ =\ 
S (t, x_i)\ ,
\label{eq:waveeq1}
\end{equation}
where the source function $S(t,x_i)$ is given by
\begin{equation}
S(t,x_i) = {\partial \over \partial x_i} \left [ c_s^2 {{\partial 
(\rho u_i)} \over \partial t} \right ] + {\partial \over \partial x_i}
\left [ c_s^2 {{\partial (\rho_0 u_i u_j)} \over {\partial x_j}} \right ] 
- {\partial \over \partial t} \left ( u_i {\partial p \over \partial 
x_i} \right ) + c_s^2 {\partial \over \partial t} \left ( \rho {\partial 
u_i \over \partial x_i} \right )   
\label{eq:source1}
\end{equation}

\noindent
It must be noted that in our model the inhomogeneous wave equations
for the wave velocity $u_i$ and density $\rho$ are of different forms.
However, there are relationships that connect all the wave variables
and they can be used to derive the wave equations for $u$ and $\rho$
once the wave equation for $p$ is known [19].

We again follow Lighthill [1] and treat the source function as being 
fully determined by a known turbulent flow.  To emphasize this point, 
we label the source function as $S_{turb} (t,x_i)$.  Since the turbulent 
flow considered in this paper is subsonic, we make a Mach-number 
expansion of the source function and retain only the lowest order terms; 
the procedure is discussed in great details by Stein [14], and it will not 
be repeated here.  This allows us to write 
\begin{equation}
S_{turb} (t,x_i) \approx {\partial \over \partial x_i}
\left [ c_s^2 {{\partial (\rho_0 u_i u_j)} \over {\partial x_j}} 
\right ]_{turb}\ .   
\label{eq:source2}
\end{equation}
which is consistent with Lighthill's results described in Sec. 2 in 
case of $c_s$ = const.

Hence, the inhomogeneous acoustic wave equation becomes 
\begin{equation}
{\partial^2 p \over \partial t^2} - c_s^2 {{\partial^2 p} \over 
{\partial x_i^2}} - 2 c_s \left ( {{\partial c_s} \over {\partial \tau_i}} 
\right ) {\partial p \over \partial x_i}\ =\ S_{turb} (t, x_i)\ ,
\label{eq:waveeq2}
\end{equation}
and 
\begin{equation}
S_{turb} (t,x_i) = c_s^2 {\partial^2 \over {\partial x_i \partial x_j}}
\left ( \rho_0 u_{ti} u_{tj} \right ) + 2 c_s \left ( {{\partial c_s} \over 
{\partial \tau_i}} \right ){\partial \over \partial x_j} \left ( \rho_0 u_{ti} 
u_{tj} \right )\ , 
\label{eq:source3}
\end{equation}
with changes in notation and replacing $[u_i]_{turb}$ in the source 
function $S_{turb}$ by $u_{ti}$. It must be noted that Eq. 
(\ref{eq:waveeq2}) reduces to Lighthill's inhomogeneous wave 
equation (see Eq. \ref{eq:LWE}) in the limit of $c_s$ = const 
and with $p = c_s^2 \rho$.

\subsection{Transformed wave equation}

To remove the nonconstant coefficient $c_s^2$ from the term with the 
second-order derivative in Eq. (\ref{eq:waveeq2}), we introduce the 
new variable $\partial \tau_i = \partial x_i / c_s$ and obtain
\begin{equation}
{\partial^2 p \over \partial t^2} - {\partial^2 p \over \partial \tau_i^2} 
- {1 \over c_s} \left ( {{\partial c_s} \over {\partial \tau_i}} \right ) 
{\partial p \over \partial \tau_i}\ =\ S_{turb} (t, \tau_i)\ ,
\label{eq:kg1}
\end{equation}
with 
\begin{equation}
S_{turb} (t,\tau_i) = {\partial^2 \over {\partial \tau_i \partial 
\tau_j}} \left ( \rho_0 u_{ti} u_{tj} \right ) + {1 \over c_s} 
\left ( {{\partial c_s} \over {\partial \tau_i}} \right ) {\partial 
\over \partial \tau_j} \left ( \rho_0 u_{ti} u_{tj} \right )\ . 
\label{eq:source4}
\end{equation}

As the next step, we remove the first order derivative from Eq. 
(\ref{eq:kg1}) by using the following transformation [20,21]:
\begin{equation}
p (t,\tau_i) = p_1 (t, \tau_i) e^{- I_c}\ ,
\label{eq:transf1}
\end{equation}
where
\begin{equation}
I_c = {1 \over 2} \int_{\tau_{i0}}^{\tau_i} {1 \over c_s} \left ( 
{{\partial c_s} \over {\partial \tilde \tau_i}} \right ) \tilde
d \tau_i\ .
\label{eq:transf2}
\end{equation}
This gives
\begin{equation}
{{\partial^2 p_1} \over {\partial t^2}} - {{\partial^2 p_1} \over 
{\partial \tau_i^2}} + \Omega_{i}^2 ( \tau_i ) p_1 = S_{turb} (t, 
\tau_i)\ ,
\label{eq:kg2}
\end{equation}
where 
\begin{equation}
\Omega_{i}^2 (\tau_i) = {1 \over 2} \left [ {1 \over c_s} \left ( 
{{\partial^2 c_s} \over {\partial \tau_i^2}} \right ) - {1 \over 
 2 c_s^2} \left ( {{\partial c_s} \over {\partial \tau_i}} \right )^2
\right ]\ , 
\label{eq:cutoff1}
\end{equation}
and
\begin{equation}
S_{turb} (t,\tau_i) = \left [ {\partial^2 \over {\partial \tau_i 
\partial \tau_j}} \left ( \rho_0 u_{ti} u_{tj} \right ) + {1 
\over c_s} \left ( {{\partial c_s} \over {\partial \tau_i}} \right ) 
{\partial \over \partial \tau_j} \left ( \rho_0 u_{ti} u_{tj} \right ) 
\right ] e^{I_c}\ , 
\label{eq:source4}
\end{equation}
\\ 
which shows that the source function is determined by both 
local ($\partial c_s / \partial x_i$) and global ($I_c$) effects.

\subsection{Solution to wave equation and acoustic cutoff frequency}

To solve Eq. (\ref{eq:kg2}), we must specify the temperature gradient
in our nonisothermal model. Since the gas pressure $p_0 = R T_0 \rho_0 
/ \mu$, where $R$ is the universal gas constant and $\mu$ is the mean 
molecular weight, must be constant in our model, the temperature and 
density variations with height must be related to each other. Let us 
consider a model in which both $T_0$ and $\rho_0$ vary only in one 
direction, say, the $x_3$ (or z) direction and assume that $T_0 (\xi) 
= T_{00} \xi^2$, where $\xi = z/z_0$, with $z_0$ being a given height, 
and $T_{00}$ is a temperature at this height. For the model to be in 
hydrostatic equilibrium ($p_0$ = const), the density $\rho_0$ must 
decrease with $\xi$ as $\rho_0 = \rho_{00} \xi^{-2}$. In this model, 
the speed of sound $c_s$ is a linear function of $\xi$ and we have 
$c_s (\xi) = c_{s0} \xi$, where $c_{s0}$ is the speed of sound at 
$z_0$. 

The nonisothermal model requires that $\tau_1 = x_1 / c_s$ and
$\tau_2 = x_2 / c_s$ but $\tau_3 = ln \vert \xi\ \vert / \omega_0$,
where $\omega_0 = c_{s0} / z_0$. Hence, we obtain $\xi = e^{\omega_0 
\tau_3}$ and $c_s (\tau_3) = c_{s0} e^{\omega_0 \tau_3}$. We use 
these results to calculate $\Omega_i^2$ (see  Eq. \ref{eq:cutoff1}), 
which gives $\Omega_1^2 = 0$, $\Omega_2^2 = 0$ and $\Omega_3^2 = 
\Omega_0^2$ with $\Omega_0^2 = \omega_0^2 / 4$ or
\begin{equation}
\Omega_0^2 = {c_{s0}^2 \over {4 z_0^2}}\ . 
\label{eq:cutoff2}
\end{equation}
%

%This allows us to write the inhomogeneous wave equation (see 
%Eq. \ref{eq:kg2}) in the following form:  
%
%\begin{equation}
%{{\partial^2 p_1} \over {\partial t^2}} - {{\partial^2 p_1} \over 
%{\partial \tau_i^2}} + \Omega_{0}^2 p_1 = S_{turb} (t, 
%\tau_i)\ ,
%\label{eq:kg3}
%\end{equation}
%
%where $\Omega_1^2 = \Omega_2^2 = 0$, $\Omega_3^2 = \Omega_0^2$ and 
%
%\begin{equation}
%S_{turb} (t,\tau_i) = \left [ {\partial^2 \over {\partial \tau_i 
%\partial \tau_j}} \left ( \rho_0 u_{ti} u_{tj} \right ) + 2 \Omega_0 
%{\partial \over \partial \tau_j} \left ( \rho_0 u_{ti} u_{tj} \right ) 
%\right ] e^{\Omega_0 \tau_3}\ . 
%\label{eq:source5}
%\end{equation}
%

An interesting result is that $\Omega_0$ is constant in the $\tau$-space
(but not in the $z$-space), so we can formally make Fourier transforms 
in time and $\tau$-space.  Based on the form of Eq. (\ref{eq:kg2}) 
and the fact that $\Omega_0$ = const, we conclude that $\Omega_0$ 
is the acoustic cutoff frequency introduced by Lamb [22,23] for acoustic 
waves propagating in a stratified but isothermal atmosphere.  The origin
and physical meaning of $\Omega_0$ is the same as the cutoff introduced
by Lamb.  The acoustic cutoff plays important roles in studying Earth's 
[24,25] and Jupiter's [26] oscillations as well as in helioseismology [27] 
and asterioseismology [28].

We make Fourier transforms in time and $\tau$-space
\begin{equation}
p_1 (t, \tau_i) = \int \int p_2 (\omega, k_i) e^{i(\omega t-k_i \tau_i)}
d \omega\ d^3k_i\ ,
\label{eq:fourier1}
\end{equation}
where $k_i$ is a wave vector corresponding to $\tau_i$. This gives
\begin{equation}
\int \int [-\omega^2 + k_i^2 + \Omega_i^2] p_2 (k_i,\omega) e^{i(\omega 
t-k_i \tau_i)}\ d\omega\ d^3k_i = S_{turb} (t, \tau_i)\ ,
\label{eq:fourier2}
\end{equation}
and
\begin{equation} 
p_2 (\omega, k_i) = {{S_{turb} (\omega, k_i)} \over {- \omega^2 + k_i^2 +
\Omega_i^2}}\ ,
\label {eq:solution}
\end{equation}
with
\begin{equation} 
S_{turb} (\omega, k_i) = {1 \over (2\pi)^4} \int \int S_{turb} (t, \tau_i) 
e^{-i(\omega t - k_i \tau_i)}\ dt\ d^3 \tau_i\ .
\label {eq:fourier4}
\end{equation}
Substituting Eq. (\ref{eq:source4}) in Eq. (\ref{eq:fourier4}) and integrating
 by parts results in:
\begin{eqnarray}
S_{turb}(\omega,k_i)={-1 \over (2\pi)^4}\int (\rho_0 u_i u_j)({\partial^2 \over 
\partial \tau_i \partial \tau_j}+{c'_{s} \over c_{s}}{\partial  \over \partial \tau_j})
e^{I_c} e^{-i(\omega t-k_i \tau_i)}d^3 \tau_i \hspace{1 mm}\ ,
\label{eq:source1}
\end{eqnarray}
where $c_s^{\prime} = d c_s /d \tau_3 = d c_s /dz$.

After some calculations (see Appendix A), $S_{turb}(t,\tau_i)$ can be written as 
\begin{equation}
S_{turb}(t, \tau_i)=e^{I_c} \rho_0 \big \{k_i k_j u_i u_j -{1 \over 4}
\left ( {c'_{s} \over c_{s}} \right )^2 u_3 u_3 - {c_{s}\prime\prime \over 2 c_{s}} 
- {i \over 2 c_{s}}[{1 \over 2}c'_{s}u_3 k_i u_i+{3 \over 2}c'_{sz}u_3 k_ju_j] \big \}\, 
\label{eq:source2}
\end{equation}
where $c_s^{\prime \prime} = d c_s^{\prime} /d \tau_3 = d c_s^{\prime} /dz$.
The above expression is the final form of the source function for the problem under 
consideration.

\section{Emitted acoustic wave energy flux}

\subsection{Mean acoustic energy flux}

\noindent 
The mean acoustic energy flux is calculated by:
\begin{equation}
\vec F=<p u^*>
\end{equation}
From the momentum conservation and continuity equation, after
ignoring gravity and non linear terms, the velocity of the fluid 
\footnote{See Eq. 54 in Stein [14]} is expressed in terms of the 
pressure perturbation, as
\begin{equation}
u_i=-{1 \over \rho_0}({\partial \over \partial t})^{-1} 
{1 \over c_s}{\partial p \over \partial \tau_i}.
\end{equation}

\noindent 
Substituting this relation into the energy flux equation leads to:
\begin{equation} \label {eq:Flux_Eq}
\vec F=-<p {1 \over \rho_0 c_s} ({\partial \over \partial t})^{-1}
 {\partial p^* \over \partial \tau_i}>
\end{equation}
\noindent 
and $p$ can be expressed in terms of its Fourier transform (see Appendix B) as
\begin{equation} \label {eq:p}
p_2(t,\tau_i)= \int {S_{turb}(\omega,k_i) \over {-\omega^2+k_i^2+\Omega_i^2}} 
e^{i\omega t-i k_i \tau_i}d^3k_i d \omega\ ,
\end{equation}
which gives the following acoustic wave energy flux 
\begin{equation} \label {eq:Flux}
\vec F(\omega,k_i)=\lim_{T \rightarrow \infty} {2 \pi \over T} {1 \over \rho_0 c_s}
\int ({k_i \over \omega}) {S_1(\omega,k_i)S^*_1(\omega,k_i) d^3 k_i d^3 k_i 
\over \big \{-{\omega}^2+{k_i}^2+\Omega_{i}^2 \big \}
\big \{-{\omega}^2+{k_i}^2+\Omega_i^2\big \}}\ ,
\end{equation}
where the integral is now evaluated.

\subsection{Asymptotic Fourier transform}
\noindent
Lighthill's formula [7] for the asymptotic value of Fourier transforms
far from the source is given by 
\begin{equation} \label {eq:Asymptotic}
\int {F(\vec k) \over G(\vec k)} e^{(i \vec k \cdot \vec \tau)}d^3 \vec k
={4 \pi^2 \over |\vec \tau|} \sum_k {F(\vec k)e^{i \vec k \cdot \vec \tau}
\over |\Delta_k G| \sqrt K}
\end{equation}
where G, the denominator in the integral in (\ref{eq:Flux}), is defined as 
$G=\vec k_i \cdot \vec k_i-(\omega^2-\Omega_i^2)$. K is the Gaussian 
curvature of the slowness surface \footnote{The slowness surface is the 
surface in wave number space where the dispersion relation is satisfied.} 
on which the direction of normal is defined as $|\hat \tau|={\Delta_k G 
\over |\Delta_k G|}={(k_1,k_2,k_3) \over \sqrt {(k^2_1+k^2_2+k^2_3)}}$. 
Refering to (\ref{eq:p}), only those wave numbers and frequencies contribute 
to the pressure perturbation where the denominator, G, vanishes, i.e, where 
the dispersion relation (G=0) is satisfied. The sum is over the set $\vec k$ 
on the slowness surface, G=0. The cosine of the angle between the sound 
propagation (group velocity), $\hat \tau$, and the vertical is $\cos \theta=
\hat z \cdot \hat \tau={k_3 \over |k|}$. The cosine of the angle between 
the wave vector, $\vec k$, and the vertical is $\cos \theta_k={{\hat z \cdot 
\vec k} \over |k|}={k_3 \over |k|}$.  Hence in our particular case the 
direction of the wave vector is the same as the direction of propagation, 
that is, $\theta_k=\theta$. 

By applying (\ref{eq:Asymptotic}) to (\ref{eq:Flux}), we obtain the 
acoustic flux at large distances given by 
\begin{equation}
\vec F(\omega,k_i)=\lim_{T \rightarrow \infty} {8 \pi^5 \over T}
{\hat \tau_i \over |\tau_i|^2} {\sqrt {(\omega^2-\Omega^2_i)} \over
\omega \rho_0 c_s} |S_{turb}(\omega,k_i)|^2\ ,
\end{equation}
which requires the source function $S_{turb}(\omega,k_i)$ to be 
determined by evaluating spectral efficiency and specifying turbulence 
and its spectrum.

\subsection{Evaluation of spectral efficiency}

\indent In the absence of generally accepted model of turbulence,
the description of turbulent flow is based on two-point, two-time
velocity correlation functions which are obtained by considering the
source at two points in a turbulent fluid at two different times [29,30].
\begin{eqnarray}
S^*_{turb}(\omega,k_i)={1 \over (2 \pi)^4} \int S^*_{turb}(t'',\tau''_i)
e^{-i(\omega t''-k_i \tau''_i)} d^3 \tau''_i dt'' \label {eq:S''}\\
S_{turb}(\omega,k_i)={1 \over (2 \pi)^4} \int S_{turb}(t',\tau'_i)
e^{i(\omega t'-k_i \tau'_i)} d^3 \tau'_i dt'\ . 
\label{eq:S'}
\end{eqnarray}
where primed and double-primed refer to the two turbulent source
points. 

Averaging the position and time, $\vec \tau_0$ being the
vector to the mean position between the two turbulent source points
and $t_0=0.5(t''+t')$ being the mean time between $t''$ and $t'$, the
coordinates in the evaluation of $|S(\vec k,\omega)|^2$ are
transformed: \setlength\arraycolsep{1pt}
\begin{eqnarray}
|S_{turb}(\omega,k_i)|^2={1 \over (2 \pi)^8}\int \int \int \int 
S^*_{turb}(t_0+{t \over 2},\tau_{0i}
+{\tau_i \over 2}) \nonumber \\ S_{turb}(t_0-{t \over 2},\tau_{0i}-
{\tau_i \over 2})
 e^{-i\omega t+i k_i \tau_i} d^3\tau_i \hspace{1 mm} d^3 \tau_{0i}
 \hspace{1 mm} dt \hspace{1 mm} dt_0 \nonumber
\end{eqnarray}
where $\tau_i=\tau''_i-\tau'_i$ and $t=t''-t'$ are the space and
time intervals between the two points respectively.  Performing 
the integration over $t_0$ will result in the time averaging of 
$|S_{turb}(\omega,k_i)|^2$:
\begin{eqnarray}
|S_{turb}(\omega,k_i)|^2={T \over (2 \pi)^8}\int \int \int \int 
<S^*_{turb}(t_0+{t \over 2},\tau_{0i}
+{\tau_i \over 2}) \nonumber \\ S_{turb}(t_0-{t \over 2},\tau_{0i}-
{\tau_i \over 2})> e^{-i\omega t+i k_i \tau_i} d^3\tau_i \hspace{1 mm}
 d^3 \tau_{0i} \hspace{1 mm} dt \label {eq:modulus}
\end{eqnarray}

Using subscript {i,j} and {l,m} for two different locations and two different times,
Eq 21 can be employed to calculate $|S^*_{turb}(t_0+{t \over 2},\tau_{0i}
+{\tau_i \over 2})S_{turb}(t_0-{t \over 2},\tau_{0i}-{\tau_i \over 2})|$.
\begin{eqnarray}
|S^*_{turb}(t_0+{t \over 2},\tau_{0i}
+{\tau_i \over 2})S_{turb}(t_0-{t \over 2},\tau_{0i}-{\tau_i \over 2})|=
(\rho^2_o e^{2I_c})[k_i k_j u'_i u'_j\nonumber\\-{1 \over 4} {(c'_{sz})^2 
\over c_{s}^{2}} u'_3 u'_3 -{c_{sz}\prime\prime \over 2 c_{s}}u_{3} 
\prime u_{3} \prime + {i \over 2 c_{s}}({3 \over 2}c'_{sz} u'_3 k_j u'_j
+{1 \over 2}c'_{sz} u'_3 k_i u'_i)] [k_l k_m u''_l u''_m-\nonumber\\-{1 
\over 4} {(c'_{sz})^2 \over c_{s}^{2}} u'_3 u'_3 -{c_{sz}\prime\prime 
\over 2 c_{s}}u_{3} \prime u_{3} \prime - {i \over 2 c_{s}}({3 \over 2}
c'_{sz} u'_3 k_j u'_j+{1 \over 2}c'_{sz} u'_3 k_i u'_i)]\ .
\nonumber\\
\label{eq:Source}
\end{eqnarray}

Spectral efficiency, $\Upsilon(t,\vec \tau)$, is obtained by expanding the above equation, 
ignoring the complex part because being an odd fuction of $\omega$ it disappears upon 
integration over $\omega$.  Then, the result is 
\begin{equation}
   \begin{aligned}
\Upsilon(t,\tau_i) & =k_ik_jk_lk_mu'_iu'_ju''_lu''_m- {1 \over 2}\left[{1 \over 2}
\left({c'_{sz} \over c_s}\right)^2+{c_{sz}\prime \prime \over c_s} \right]
k_ik_ju'_iu'_ju''_3u''_3\\
&- {1 \over 2}\left[{1 \over 2}\left({c'_{sz} \over c_s}\right)^2+{c_{sz}
\prime \prime \over c_s} \right]k_l k_m u''_l u''_mu'_3u'_3\\
&+ {1 \over 4}\left[{1 \over 4}\left({c'_{sz} \over c_s}\right)^4+
\left ({c_{sz}\prime \prime \over c_s} \right)^2+{(c'_{sz})^2 c_{sz}\prime 
\prime \over c_{s}^3}\right] u'_3u'_3u''_3u''_3\\
&+\left({c'_{sz}\over c_{s}}\right)^2 k_ju'_jk_m u''_m u'_3 u''_3\ ,
   \end{aligned}
\label{eq:Upsilon}
\end{equation}
\noindent
which allows writing the source function as 
\begin{equation}
|S_{turb}(\omega,k_i)|^2={T e^{2I_c} \over (2 \pi)^8} \int \int \int \rho_0^2 
\Upsilon(t,\tau_i) e^{-i\omega t+i k_i \tau_i} d^3 \tau_i d^3 \tau_{0i} dt\ ,
\end{equation}
\noindent
and the acoustic wave energy flux becomes 
\begin{equation}
\vec F(\omega,k_i)={2\pi \hat \tau_i \over |\tau_i|^2}{\sqrt {(\omega^2-\Omega_i^2)} 
\over \omega}\int \int \int {\rho_0(\tau_{0i}) \over (4 \pi^2)^2 c_s(\tau_{0i})}
\Upsilon(\vec \tau,t)e^{-i\omega t+i k_i \tau_i} d^3 \tau_i d^3\tau_{0i} dt\ .
\end{equation}

\setlength\arraycolsep{1pt}
Substituting the value of $c'_{sz}/c_{s}=2\Omega_0$, Eq. \ref{eq:Upsilon} simplifies to
\begin{eqnarray}
\Upsilon(t,\tau_i)=<(k_i u'_i)^2(k_i u''_i)^2>-3\Omega_0^2 \{<(k_i u'_i)^2 
u''_3 u''_3>+ <(k_i u''_i)^2 u'_3 u'_3>\} \nonumber\\+ 9 \Omega_0^4<u'_3 
u'_3 u''_3 u''_3> + 4 \Omega_0^2 <(k_i u'_i)(k_i u''_i)u'_3 u''_3>\ . 
\label{eq:Upsilon_i}
\end{eqnarray}

\noindent 
The fourth-order velocity correlation in Eq. (\ref{eq:Upsilon_i}) can be reduced
to a second-order velocity correlation [30], and the result is 
\setlength\arraycolsep{1pt}
\begin{equation}
<u_1 u_2 u'_3 u'_4>=<u_1 u_2><u'_3 u'_4>+<u_1 u'_3><u_2 u'_4>+
<u_1 u'_4><u_2 u'_3>\ , 
\label{eq:reduced}\\
\end{equation}
which gives
\begin{eqnarray}
\Upsilon(t,\tau_i)=2w^4 <v' v''>^2+ 4 \Omega_0^2 w^2 <v'v''><u'_3
u''_3> - 8\Omega_0^2 w^2<v' u''_3>^2\nonumber\\+
18 \Omega_0^4<u'_3 u''_3>^2 \label {eq:Upsilon_f}\ ,
\end{eqnarray}
that is derived in Appendix C.

\subsection{Convolution of turbulence energy spectra}

\noindent Generally, the Fourier transform of the product of the
second-order velocity correlations is expressed as
\begin{eqnarray}
{1 \over (2 \pi)^4}\int d^3\vec \tau \int_{-\infty}^{\infty} e^{-i
(\omega t- \vec k \cdot \vec \tau)} dt <u'_l u''_m><u'_n u''_o>
\nonumber \\ =\int \int \lambda_{lm}(\vec k-\vec
p,\omega-\sigma)\lambda_{no}(\vec p,\sigma)d^3\vec p \hspace{1 mm}
d\sigma=J_{lmno}\ , 
\label {eq:J}
\end{eqnarray}
where $\lambda_{ij}$, the Fourier transform of the velocity correlation 
$<u'_i u''_j>$, is defined as:
\begin{equation}
\lambda_{ij}={1 \over (2 \pi)^4}\int \int <u_i(\vec x,t_0)
u_j(\vec x+\vec r,t_0+t)> e^{i (\omega t- \vec k \cdot \vec r)}
d^3 \vec r \hspace{1 mm} dt\ .
\end{equation}

\indent From the phenomenological treatment of turbulence, the
correlations between the instanteous velocity components at two
different locations in the turbulent region can be evaluated when 
a turbulent energy spectrum $E(\vec k,\omega)$ is specified. 
Assuming the turbulence to be isotropic, homogeneous and 
incompressible [29,30], $\lambda_{ij}$ can be expressed as
\begin{equation}
\lambda_{ij}(\vec k,\omega)={E(\vec k,\omega) \over {4 \pi k^2}}
(\delta_{ij}-{k_i k_j \over k^2})\ .
\label{eq:lambda}
\end{equation}

\indent Even though the medium is nonisothermal, it can be treated
locally as homogeneous and isotropic, hence the application of the
above equation. It is further assumed that the turbulence energy
spectrum, $E(\vec k,\omega)$, can be factored into the frequency
independent spatial turbulent energy spectrum E(k) and the turbulent
frequency factor $\Delta(\omega,k)$, $E(\vec
k,\omega)=E(k)\Delta(\omega,k)$, which in turn can be substituted in
(\ref{eq:lambda}) to simplify the calculation of (\ref{eq:J}).
\begin{eqnarray}
{1 \over (4 \pi)^2}\int \int {E(\vec k-\vec p) \over q^2} {E(\vec p)
\over p^2} \Delta (\omega-\sigma,\vec k-\vec p) \Delta (\sigma,\vec p)
\nonumber \\ (\delta_{lm}-{k_l k_m \over q^2})(\delta_{no}-{k_n k_o 
\over p^2}) d^3\vec p \hspace{1 mm} d\sigma=J_{lmno}\ ,
\label {eq:J_{lmno}}
\end{eqnarray}
\noindent
where $\vec q \equiv \vec k-\vec p$. The integration of $\sigma$, 
which has the frequency terms only, can be performed separately,
$$g(p,q,\omega)\equiv \int_{-\infty}^{\infty} \Delta (\omega-\sigma,\vec q) 
\Delta (\sigma,\vec p) d \sigma$$ and Eq. (\ref{eq:J_{lmno}}) can 
be rewritten as
\begin{eqnarray}
{1 \over (4 \pi)^2}\int {E(\vec q) \over q^2} {E(\vec p) \over p^2} 
g(p,q,\omega) (\delta_{lm}-{k_l k_m \over q^2})(\delta_{no}-
{k_n k_o \over p^2})d^3p=J_{lmno}\ .
\end{eqnarray}

The integration of $d^3 \vec p$ is simplified by taking $\vec k$ as
the axis of the spherical cordinate system, $d^3p=p^2 dp \sin \theta
d\theta d\phi=p^2 dp d\mu d\phi=2 \pi p^2 dp d\mu$ with $\mu=\cos
\theta_{pk}=\cos \theta$ and $|q|=\sqrt {(k^2+p^2-2 kp\mu)}$. 
The Fourier transforms of velocity correlations appearing in (\ref{eq:Upsilon_f}) 
contain four terms, namely $J_{kkkk},J_{kzkz},J_{kkzz}$ and $J_{zzzz}$. Next
we substitue $J_{kkkk},J_{kzkz},J_{kkzz}$ and $J_{zzzz}$ in
(\ref{eq:Upsilon_f}), take $({1 \over 8 \pi}) \int_{0}^{\infty}dp
\int_{-1}^{+1} d\mu {E(q) E(p) \over q^2} g(p,q,\omega)$ as common,
and define the remaining equation as $f(\omega,\theta,p,q,\mu)$.
$$f(\omega,\theta,p,q,\mu)=f_q+f_d+f_m$$
$$f_q=2(w)^4{p^2 \over q^2}(1-\mu^2)^2\textrm{,}$$
\begin{eqnarray}
f_d=-8w^2 \Omega^2_0\bigg \{\mu^2 \cos^2 \theta_k (1-{p^2 \over
q^2}(1-\mu^2))+({p^2 \over q^2}\mu^2-{pk\mu \over q^2}){1 \over
2}(1-\mu^2)\sin^2 \theta_k \bigg \}\, \nonumber
\\+4w^2\Omega^2_0\bigg \{{p^2 \over q^2}(1-\mu^2)\{1-\mu^2 \cos^2
\theta_k-{1 \over 2}(1-\mu^2) \sin^2
\theta_k\}\bigg\}\textrm{,}\nonumber
\end{eqnarray}
and
\begin{equation}
   \begin{aligned}
f_m &=18 \Omega^4_0\Bigg \{1-\mu^2 \cos^2 \theta_k 
-{1 \over 2}(1-\mu^2) \sin^2 \theta_k+{k^2 \over q^2}
\{-\cos^2 \theta_k+\mu^2 \cos^4 \theta_k\nonumber\\ 
&+{1\over 2}(1-\mu^2)\cos^2 \theta_k \sin^2 \theta_k \}+{p^2
\over q^2}\{-\mu^2 \cos^2 \theta_k+\mu^4 \cos^4 \theta_k 
\nonumber\\
&+3\mu^2(1-\mu^2) \cos^2 \theta_k \sin^2 \theta_k 
-{1 \over2}(1-\mu^2)\sin^2 \theta_k+{3 \over 8}(1-\mu^2)^2 
\sin^4\theta_k\}\nonumber\\
&+{2pk \mu \over q^2}\{\cos^2 \theta_k-\mu^2 
\cos^4\theta_k-{3 \over 2}(1-\mu^2)\cos^2\theta_k \sin^2 
\theta_k\}\Bigg \}
    \end{aligned}
\end{equation}
The presented explict forms of $f(\omega,\theta,p,q,\mu)$
allow us to calculate and give the final expression for the 
acoustic wave energy flux generated by turbulent motions.

\subsection{Acoustic wave energy flux and its discussion}

\noindent
The emitted acoustic energy flux for a given frequency,
calculated in $\tau$ space, is given by
\begin{eqnarray}
\vec F(t,\tau_i)={1 \over 16} {\hat \tau_i \over |\tau_i|^2}
{(\omega^2-\Omega^2_i)^{1/2} \over \omega} e^{2I_c} 
\int {\rho_0(\tau_{0i}) \over c_s(\tau_{0i})} d^3\tau_{0i} 
\nonumber \\ 
\int_{0}^{\infty}dp \int_{-1}^{+1} d\mu {E(q) E(p) \over 
q^2} g(p,q,\omega) f(\omega,k,p,q,\mu)\ ,
\end{eqnarray}
and the expression is valid for a medium with temperature
and density gradients related to each other, and with constant 
pressure, which means that there is no gravity.  The flux was 
derived under the assumption of isotropic and homogeneous 
turbulence, whose spectrum and frequency factor must be 
specified. 

As the above results show, the generated acoustic wave energy 
flux is calculated by making the multipole expansion of the source 
function given by Eq. (\ref{eq:Source}).  This allows identifying
contributions from different wave sources and writing the source 
function in the following form: $S_a [ p_o , u_t ] = S_{quadrupole} 
+ S_{dipole} + S_{monopole}$, where $S_{quadrupole} \sim 
\omega^4$, $S_{dipole} \sim \omega^2 \Omega_0^2$ and 
$S_{monopole} \sim \Omega_0^4$.  The dipole and monopole 
source terms describe the conversion of kinetic energy into acoustic 
energy resulting from forcing the mass and momentum in a fixed 
region of space to fluctuate, respectively, and are produced by the 
density gradient.  

In the original approach presented by Stein [14], see also [16], the 
external gravitational force is responsible for the density gradient 
and stratification of the background solar atmopshere, whcih is 
assumed to be isothermal.  Studies by Goldreich and Kumar [18]
revealed that the forced turbulence occuring in the solar convection 
zone is driven by the fluctuating buoyancy force, which represents 
the coupling of gravity to density fluctuations associated with the 
turbulent field, and leads to dipole emission that may dominate 
over the quadrupole source.  There are similarities between the 
dipole terms obtained in these two paper, which shows that the 
fluctuacting buoyancy force can be accounted for by either the 
method proposed in [18] or by the multipole expansion considered
in [14].

In all previous approached to the acoustic wave generation 
by isotropic and homogeneous turbulence [1,2,6,14,16,18],
the background medium was assumed to be isothermal. The 
results presented in this paper demonstrate that the temperature
gradient lead to monopole and dipole emissions, which depend
directly on the acoustic cutoff frequency.  The model considered
in this paper is simple, nevertheless, its results are important 
because temperature gradients exist in realistic planetary, solar 
and stellar atmopsheres, and they affect the hydrostatic equilibrium
of such atmospheres by changing their denisties and pressures.
Moreover, temperature gradients also make the wave speed to 
be a function of atmopsheric height, which modifies the wave
cutoff frequencies, and wave propagation conditions.  

The acoustic power output, $P_s$, obtained in this paper can 
be written in the following form
\begin{equation} 
P_s = {{\rho_o u_t^3} \over {l_o}} \left ( \alpha_q M^5 +
\alpha_d M^3 + \alpha_m M \right )
\end{equation}
where $\alpha_q$, $\alpha_d$ and $\alpha_m$ represent 
the emissivity coefficients of quadrupole, dipole and monopole
sources, respectively, and $M = u_t / c_s$ is the turbulent 
Mach number.  The previous results (e.g., [1,2,14]) showed 
that quadrupole emission dominates in the acoustic wave 
energy spectrum, and that it is a sensitive function of the 
wave frequency, the turbulent energy spectrum, the turbulent 
frequency factor, and the physical parameters in the region 
of wave generation [16].  However, there are some special 
conditions when contributions from the dipole and monopole 
sources may become important (e.g., [13,18]).

The results presented in this paper can be used to include 
tempearature gradients in Stein's approach [14] that 
considered only stratification.  Such an extension would 
make the theory of sound generation to be applicable 
to variuos background media with different gradients in 
physical parameters, including planetary, solar and stellar 
atmospheres.  However, based on the theory developed 
in this paper, it seems unlikely that it can be done analytically;
instead, it would require numerical simulations (e.g., [31]), 
similar to those described in [32] or recently performed 
in [33]. 

It must be also pointed out that the presence of the acoustic
cutoff frequency in the the developed theory of sound generation
restricts frequencies the genearted waves may have to those that 
are higher than the cutoff at the wave source.  Moreover, the 
wave propagation away from the source is affected by the 
cuttoff and may lead to wave reflection that may limit the 
wave energy transfer to the layers located above the wave 
source.  Both effects make the presented theory more realistic 
than the original Lighthill's theory of acoustic wave generation.

\section{Conclusions}

The theory of generation of acoustic waves by a region of isotropic 
and homogeneous turbulence embedded in a medium with gradients 
in temperature and density, but constant pressure, is developed.
The theory extends the original Lighthill theory of sound generation 
by accounting for the effects of the gradients on the wave source,
and on the wave propagation outside the source.  It is shown that 
the gradients lead to monopole and dipole wave emission, whose
efficiency may exceed Lighthill's quadrupole emission depending 
on the values of the gradients.  The wave propagation away from 
the source is also affected by the gradients, which give the origin 
to the acoustic cutoff frequency.  It is the latter that uniquely 
determines the efficiency of monopole and dipole emissions, 
and also sets up the wave propagation conditions in the medium.   

The main result of this paper is the general formula for the 
generated acoustic wave energy flux.  To guarantee analytical 
solutions, the formula is obtained for a simple model of 
non-isothermal medium, in which the speed of sound varies 
linearly with height, which causes variations of density; 
however, pressure remains constant as there is no gravity 
in this model.  The presented theory and its results are 
compared to the original Lighthill work [1] as well as to 
the extended version of Lighthill's theory that was done 
by Stein [14], who showed that atmospheric stratification
is responsible for the origin of monopole and dipole wave
emissions.   The results of this paper demonstrate that 
temperature gradients are also responsible for both 
monopole and dipole emissions because they modify 
the equilibrium at the background medium; the effect
would be even more prominent in planetary, solar and 
stellar atmospheres, where temperature gradients can 
be strong and they will significantly influence the 
hydrostatic equailibrium in these atmopsheres.\\
\section{Data Availibility} 
Data sharing not applicable to this article as no datasets were generated or analysed
during the current study.\\
\section{Declaration}
Conflict of interest: On behalf of all authors, the corresponding author states that there is no conflict of interest. \\

\begin{appendices}

\section{Derivation of the source function}\label{secA}

The general form of the source function is given by Eq. 
(\ref{eq:source1}) and it can be written in the following form

\begin{eqnarray}\label{eq:s_turb}
S_{turb}(\omega,k_i)&=&{-1 \over (2\pi)^4}\int (\rho_0 u_i u_j) 
({\partial^2 \over \partial \tau_i \partial \tau_j}+c'_{si}
{\partial  \over \partial \tau_j})e^{I_c} e^{-i(\omega t-k_i \tau_i)}d^3
 \tau_i \hspace{1 mm} dt \nonumber\\
&=&{-1 \over (2\pi)^4}\int (\rho_0 u_i u_j) f(\tau_i,t)d^3 
\tau_i \hspace{1 mm} dt\ ,
\end{eqnarray}
where
\begin{eqnarray} 
f(t,\tau_i)&=&({\partial^2 \over \partial \tau_i \partial \tau_j}+
c'_{si}{\partial  \over \partial \tau_j})e^{I_c}
 e^{-i(\omega t-k_i \tau_i)}\nonumber\\
&=&[-k_i k_j+i {1 \over 2} k_j c'_{si}+i {1 \over 2} k_i c'_{sj}+{1 \over 4}
 c'_{si} c'_{sj}+c'_{si}i k_j+{1 \over 2}c'_{si} c'_{sj}]e^{I_c} 
 e^{-i(\omega t-k_i \tau_i)}\nonumber\\
&=&[-k_i k_j +{3 \over 4}c'_{si} c'_{sj}+i({1 \over 2}k_i c'_{sj}+
{3 \over 2}k_j c'_{si})]e^{I_c} e^{-i(\omega t-k_i \tau_i)}\ .
\label{eq:f}
\end{eqnarray}
Substituting Eq. (\ref{eq:f}) into Eq. (\ref{eq:s_turb}) gives
\begin{eqnarray}
S_{turb}(\omega,k_i)={1 \over (2\pi)^4}\int \big \{k_i k_j -{3\over 4}
c'_{si}c'_{sj}-i[{1 \over 2}c'_{sj}k_i+{3 \over 2}c'_{si}k_j] \big \}
(e^{I_c} \rho_0 u_i u_j) \nonumber\\ e^{-i(\omega t-k_i \tau_i)}
d^3 \tau_i \hspace{1 mm} dt\ ,
\end{eqnarray}
which when compared to 
\begin{eqnarray}
S_{turb}(\omega,k_i)&={1 \over (2\pi)^4} \int S_{turb}(t, \tau_i)
 e^{-i(\omega t-k_i \tau_i)} d^3 \tau_i \hspace{1 mm} dt\ ,
\end{eqnarray}
results in
\begin{equation}
S_{turb}(t,\tau_i)=\big \{k_i k_j -{3 \over 4}c'_{si}c'_{sj}-i
[{1 \over 2}c'_{sj}k_i+{3 \over 2}c'_{si}k_j]
 \big \}(e^{I_c} \rho_0 u_i u_j)\ ,
\end{equation}
that is Eq. (\ref{eq:source2}) in the main text.\\

\section{Calculation of the acoustic flux}\label{secB}

\noindent Time averaging of the flux requires the frequency of both
factors, $\vec p$ and $\vec u$, to be the same. Hence
$k'_i=k''_i=k_i$. Substituting (\ref{eq:p}) into (\ref{eq:Flux_Eq})
and performing the required algebra, the Flux turns out to be:
\begin{equation}
\vec F(k_i)=\lim_{T \rightarrow \infty} {1 \over T} \int_{-T/2}^{T/2} dt 
{1 \over {\rho_0 c_s}} {\int {k_i \over \omega''}{S_{turb}(\omega',k_i)
S^*_1(\omega'',k_i) e^{i(\omega'-\omega'')t} \hspace {1 mm} d^6 k_i d
\omega' d\omega'' \over \big \{-{\omega'}^2+{k_i}^2+\Omega_i^2\big \}
\big \{-{\omega''}^2+{k_i}^2+\Omega_i^2\big \}}}\ ,
\end{equation}
\noindent
where the "*" denotes a conjugate. Using $\int_{-\infty}^{\infty} 
e^{i(\omega'-\omega'')t}dt = 2 \pi \delta(\omega'-\omega'')$ to 
integrate by time, the equation is reduced to
\begin{equation}
\vec F(k_i)=\lim_{T \rightarrow \infty} {2 \pi \over T} {1 \over 
\rho_0 c_s} \int ({k_i \over \omega''}) {S_{turb}(\omega',k_i)
S^*_{turb}(\omega'',k_i) \delta (\omega'-\omega'') d^6k_i 
\hspace {1 mm} d\omega' d\omega'' \over \big \{-{\omega'}^2
+{k_i}^2+\Omega_i^2\big \}\big \{-{\omega''}^2+{k_i}^2
+\Omega_{0}^2 \big \}}\ .
\end{equation}

\noindent
Integrating by $d\omega''$ and using $\vec F(k_i)=\int 
\vec F(\omega',k_i)d \omega'$, the expression for $\vec F(\tau_i)$ 
is evaluated as
\begin{equation} 
\vec F(\omega,k_i)=\lim_{T \rightarrow \infty} {2 \pi \over T}
{1 \over \rho_0 c_s} \int ({k_i \over \omega})
{S_{turb}(\omega,k_i)S^*_{turb}(\omega,k_i) d^3 k_i d^3 k_i 
\over \big \{-{\omega}^2+{k_i}^2+\Omega_{0}^2 \big \}
\big \{-{\omega}^2+{k_i}^2+\Omega_i^2\big \}}\ ,
\end{equation}
which is Eq. (\ref{eq:Flux}) in the main text. \\

\section{Fourth-order turbulent correlations}\label{secC}

The fourth-order velocity correlation in Eq. (\ref{eq:Upsilon_i}) 
can be reduced to a second-order velocity correlation [30], which
gives
\begin{eqnarray}
\Upsilon(\tau_i,t)=w^4 <v'^2 v''^2>-3\Omega_0^2 w^2 <v'^2
{u''_3}^2>-3\Omega_0^2 w^2<v''^2 {u'_3}^2>\nonumber\\+9
\Omega_0^4<{u'_3}^2 {u''_3}^2>+16 \Omega_0^2 w^2 <v' v''
u'_3 u''_3>\ .
\end{eqnarray}
Each term in the above equation can be simplified by using 
\begin{eqnarray}
<v'v'v''v''>=2<v'v''>^2\nonumber\\ 
<v'v'u''_3 u''_3>=2<v'u''_3>^2\nonumber\\
<v''v''u'_3u'_3>=2<v''u'_3>^2\nonumber\\
<u'_3u'_3u''_3u''_3>=2<u'_3u''_3>^2\nonumber\\
<v'v''u'_3u''_3>=<v'v''><u'_3u''_3>+<v'u''_3>^2\nonumber
\end{eqnarray}
The resulting spectral efficiency turns out to be (\ref{eq:Upsilon_f}) 
in the main text after taking $<v'u''_3>^2=<v''u'_3>^2$.\\

\end{appendices}

\end{document}